# Cathode Side Transport Phenomena Investigation and Multi-Objective Optimization of a Tapered Parallel Flow Field PEMFC

**Mehrdad Ghasabehi [a], Ali Jabbary [b], Mehrzad Shams[a],***

[a] Multiphase Flow Lab, Faculty of Mechanical Engineering, K.N. Toosi University of Technology, Tehran, Iran

[b] Mechanical Engineering Department, Urmia University, Urmia, Iran

## Abstract

A Proton Exchange Membrane Fuel Cell (PEMFC) provides stable, emission-free, high-efficiency power. Water management and durability of PEMFCs are directly affected by transport phenomena at the cathode side. In the present study, transport phenomena are investigated and optimized in a tapered parallel flow field. Main channels in the flow field are tapered, which increases limiting current density by 41%. Two objectives, i.e. water saturation and transport resistance, are considered metrics for transport phenomena in a tapered parallel flow field PEMFC. Operating pressure, temperature, stoichiometries at both sides, and the porosity of gas diffusion layers are selected as parameters to be optimized. Two functions are generated for objectives by integrating 3D multiphase-flow computational fluid dynamics and Response Surface Methodology. Multi-Objective Optimization (MOO) is carried out with two different methods. Multi-Objective Particle Swarm Optimization (MOPSO) and Non-dominated Sorting Genetic Algorithm II (NSGA-II) are employed to produce two challenging Pareto fronts. The results demonstrate that MOPSO performs better than NSGA-II. MOPSO recognized quite the same Pareto front with lower runtime. In the last step, the Technique of Order Preference Similarity to the Ideal Solution (TOPSIS) is used to select an optimum point from the Pareto front. The results are compared against experimental data, and good correspondence is observed. The optimum features are temperature 323, pressure 1 atm, anode stoichiometry 3, cathode stoichiometry 2.62, and porosity 0.68. The porosity and pressure played the most significant roles in determining water saturation and resistance.

---

* Corresponding author: shams@kntu.ac.ir, Telefax: +982188677272, Postal address: No. 17, Pardis St., Mollasadra Ave., Vanak Sq., Tehran 19395-1999, Iran

## 1- Introduction

A myriad of problems and imminent threats are the repercussions of fossil fuel consumption; in recent decades, the methods of rapidly reducing greenhouse gas emissions have been a point of discussion. Also, in long-term decisions, humans should be conscious of the rapid depletion of the limited reservoir of fossil fuels. There are clean sources like hydrogen which can be our future reservoir of energy, and reliable technologies, for instance, fuel cells, especially Proton Exchange Membrane Fuel Cells (PEMFCs). PEMFCs, directly convert the chemical energy of fuel and oxidant to electrical energy [1]. They have suitable efficiency, high power density, and approximately zero emissions. Moreover, they can operate at low temperatures and pressures. However, some technical problems have constrained the complete commercialization of PEMFCs.

Several studies have been conducted to solve the challenges of commercializing PEMFCs. Thermal management [2], water management [3], and oxygen transport resistance [4] account for the majority of investigations. The operating conditions of PEMFC [5], its flow field [6], and the optimization of properties of cells [5] have all been investigated. Modified catalysts [7], the properties of the Gas Diffusion Layer (GDL) [8], the clamping pressure of cells [9], degradation [10], hydrogen economy [11] were also studied. The investigations unveiled the behaviour of PEMFC at different conditions.

In high current densities, PEMFCs suffer from concentration loss or, in other words, oxygen starvation [12]. Some studies focused on the transport of oxygen [4,12–18]. Reshetenko and Polevaya [4] investigated a single cell open flow field PEMFC and reported the resistance against oxygen transport compared to a serpentine flow field PEMFC. The single-cell open flow field has lower transport resistance at oversaturated conditions. The resistance in GDL of a PEMFC was studied by Owejan et al. [13]. They employed neutron imaging in limiting current density and showed the relation between water saturation and effective diffusion coefficient. They modified the Bruggeman relationship for two commercially available gas diffusion layer materials. Reshetenko and Ben [14] studied the impact of GDL structure on oxygen mass transport. They reported that a microporous layer added to GDL does not enhance mass transport but water management. Baker et al. [15] employed a limiting current density method. They unveiled the oxygen transport resistance in a multi-channel serpentine PEMFC. They described pressure-dependent and pressure-independent resistances and reported effective diffusion coefficients for plain Toray papers. The resistance in a parallel flow field PEMFC was studied by Wang et al. [16]. They showed the influence of operating conditions on different kinds of diffusion resistance. When the humidity increased, oxygen transport resistance increased, and the resistance was largely determined by the change of resistance in phosphoric acid. Oh et al. [17] observed oxygen mass transport in different layers of a PEMFC. They showed the influence of relative humidity on the resistance and illustrated the dominant diffusion mechanism.

GDL is a significant component of PEMFCs, and their properties affect their performance [8,19,20]. Zhang and Shi [21] investigated the impact of external factors. They reported that the porosity of GDL is the most important factor for water and oxygen distribution. Son et al. [8] analyzed the impact of directional permeability on the performance of three PEMFCs with parallel, interdigitated, and serpentine flow fields. The parallel flow field showed the lowest dependence on the direction of permeability. Yue et al. [19] studied the impact of hydrophobicity of GDL on the distribution of produced water. It was observed that implementing a hybrid hydrophobic area of GDL under channels enhances water management. The effect of GDL structure on water management was investigated by Shangguan et al. [22]. They reconstructed the geometry o GDL based on the stochastic parameter method. They showed that the distribution of GDL porosity and its contact angle significantly changed the produced water distribution. The GDL region with high porosity tends to accumulate liquid water. The hydrophilic GDL has a higher water saturation.

Operating conditions can change the performance and lifetime of a PEMFC [5]. Some researchers dedicated noticeable investigations about the impact of operating conditions on performance [23–31]. Zhang et al. [23] studied the influence of cathode side stoichiometry, pressure, and relative humidity on performance; a higher mass flow rate of air and more back pressure at saturated conditions improve the performance. Li et al. [24] investigated the performance of a PEMFC with a thin membrane. It was demonstrated how oxygen transport resistance, pressure, and relative humidity affect performance. It was shown that in a thin membrane, hydrogen crossover increases. The influence of pressure, temperature, relative humidity, and the porosity of the GDL on the performance of a single channel PEMFC was studied by Jin et al. [25]. Two Correlations for oxygen molar concentration and current density were presented, and it was declared that pressure is the most pivotal factor in performance. The impact of temperature, relative humidity, pressure, and depth and width of the channel of a parallel cathode flow field PEMFC were investigated by Carcadea et al. [26]. They showed that low relative humidity and high temperature and pressure improve cell performance, and smaller channels have superior performance. Santarelli and Torchio [27] studied the influence of relative humidity, temperature, and pressure on performance. Their findings showed that reducing relative humidity in low voltages is better, and pressure's effect at dry inlets is lower. In an investigation, Chen et al. [29] studied the influence of operating conditions on the performance of a multi-channel serpentine PEMFC. The effect of pressure, relative humidity, and stoichiometry at both sides and temperature on oxygen distribution and output voltage were reported. Yang et al. [31] investigated the cold start of a PEMFC. They revealed the importance of voltage, pressure, and inlet flow rates on the cold-start properties of the PEMFC. Ghasabehi et al. [6] proposed a modified tapered parallel flow field PEMFC and investigated the impact of operating conditions on performance. It was demonstrated that Cathode side stoichiometry and then pressure play the most critical role in the output power density of the PEMFC.

The optimization of PEMFC parameters is another subject attracting many researchers' attention. Some of the decision variables selected frequently are the ratio of the rib to channel width [32], the ratio of channel height to channel width [33], cost [34], and efficiency [34]. There are different methods for optimization with varying accuracy and applications.

The employment of optimization algorithms and artificial intelligence constitutes a noticeable number of studies related to PEMFCs [5,32,34–44]. To introduce the optimum operating conditions of an enhanced parallel flow field PEMFC, Ghasabehi et al. [5] studied the impact of operating conditions on performance. They employed Response Surface Methodology (RSM) to present two equations describing output voltage and amount of maximum power density. Then, they revealed the ideal state with a suitable efficiency, power density, and lifetime. The ideal conditions are 3.35 atm, 328 K, saturated inlets, cathode stoichiometry 4.87, and anode stoichiometry 4.23. Likewise, Li et al. [35] optimized the operating conditions and components size of a single straight channel PEMFC with three objectives named efficiency, oxygen distribution uniformity, and power density. Their reported ideal factors are 2.51 atm, 350.7 K, saturated inlets, GDL thickness 0.41 mm, membrane width 0.081mm, channel width 1.6 mm, and anode stoichiometry 1.5. Chen et al. [39] investigated the role of operating conditions on the voltage consistency of a PEMFC stack and compared different machine learning methods to forecast the stack voltage consistency. An integrated regression method was the ideal method. Li and Eghbalian [40] provided a modified flow field. They showed the application of the Black Widow Optimization Algorithm to catch the constants of an equation describing the cell's output voltage. There are different algorithms used to predict the output voltage of PEMFCs, such as Adaptive Sparrow Search Algorithm [41], Salp Swarm Algorithm [42], Sine-Cosine Crow Search Algorithm [45], Particle Swarm Optimization [45], Whale Optimization Algorithm [45], Hybrid Adaptive Differential Evolution Procedure [46], Grasshopper Optimization Algorithm [46], Particle Swarm Algorithm [38], Swarm Dolphin Algorithm [47], Grey Wolf Optimization Algorithm [48], Crow Search Algorithm [38], Teacher Learning Algorithm [38]. The accuracy and runtime ratio of different optimization methods can be different. Some investigated [45,46] compared optimization methods in specific cases. The pros and cons of various optimization models are weighted differently in different situations. A further comparison of the multi-objective version of the model for PEMFC with a variety of features and objectives will be helpful.

To the best of the authors' knowledge, there is a vacancy of straightforward, reliable relations describing transport phenomena, especially in a tapered parallel flow field PEMFC. In this study, the resistance against oxygen transport and water saturation is selected as objectives to examine transport phenomena in the PEMFC. Two functions are presented for them, which are valuable for comprehensive studies about the cell's behaviour. Pressure, temperature, stoichiometries of both sides, and the porosity of GDL are selected as the most significant factors affecting water management and resistance. The effects of selected factors on cell behaviour are studied. A comprehensive investigation is carried out by combining Computational Fluid Dynamics (CFD), RSM, and Multi-Objective Optimization (MOO) methods to present optimum features. Two MOO methods named Non-dominated Sorting Genetic Algorithm II (NSGA-II) and Multi-Objective Particle Swarm Optimization (MOPSO) are compared.

## 2- Computational domain

The modelled PEMFC contains nine parts, a membrane, two GDLs, two CLs, two flow fields, and two current collectors. Figure 1 shows all computational domains. GDL, CL, and membrane thicknesses are 0.26 mm, 0.01 mm, and 0.03 mm, respectively. The depth and width of the flow field are 1.2 mm and 1.52 mm. The flow field is an enhanced parallel flow field with 21 sub-channels. The main channels are tapered to unify oxidant distribution over the reaction site. The start width of the main channel is three-time larger than the end. The details of the flow field properties are tabulated in Table 1.

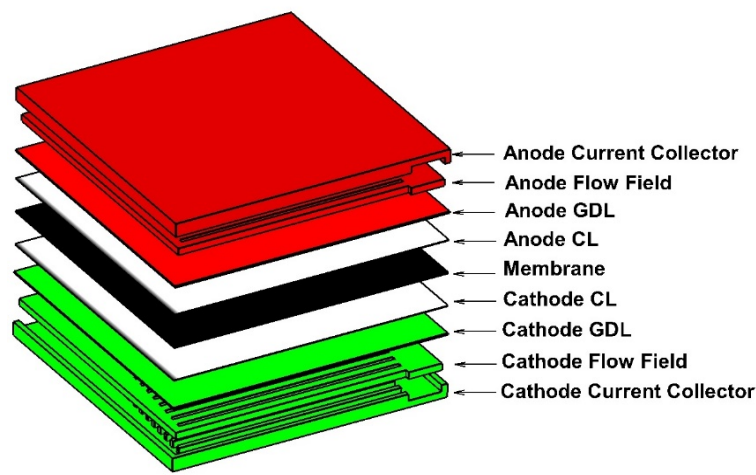

(a)

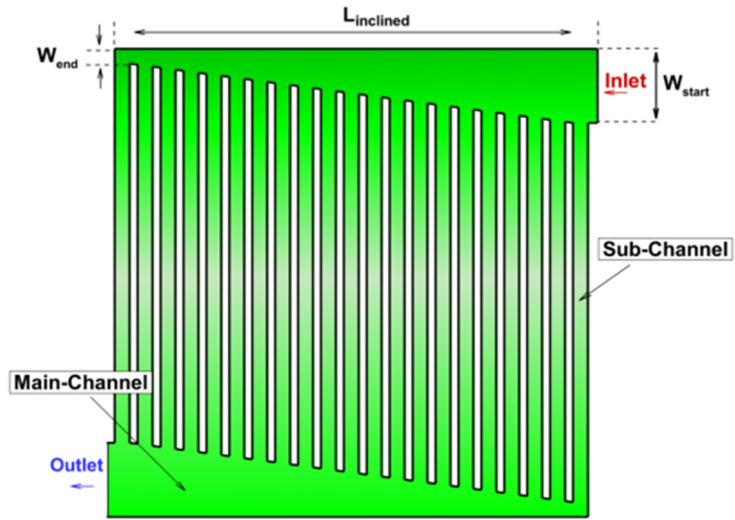

(b)

**Figure 1-** Schematic view of (a) computational domains (b) flow field

**Table 1-** The dimensions of flow field

| Parameter | Unit | Value |
|---|---|---|
| **Channel height** | mm | 1.2 |
| **Channel width** | mm | 1.52 |
| **Rib width** | mm | 0.83 |
| **BP thickness** | mm | 2.4 |
| **$L_{inclined}$** | mm | 45.48 |
| **$W_{start}$ (in tapered flow field)** | mm | 6.08 |
| **$W_{start}$ (in simple flow field)** | mm | 1.52 |
| **$W_{end}$** | mm | 1.52 |

## 3- Governing Equations

The equations modelling the transport phenomena of PEMFC in the CFD section are shown in Table 2. The equations are continuity, momentum, species transport, energy, Butler–Volmer, current transport, and the emergence and transport of liquid water.

**Table 2-** Governing equations

| Equation | | | Sources terms | |
|---|---|---|---|---|
| | | | Non zero Region | Amount |
| Continuity equation | | $\nabla \cdot (\varepsilon \rho \vec{u}) = S_m$ <br> $\rho_{mix} = \dfrac{P}{RT \sum \dfrac{Y_i}{M_i}}$ | Anode side catalyst layer | $S_{m.a} = -\dfrac{j_a}{2F} M_{H_2}$ |
| | | | Cathode side catalyst layer | $S_{m.c} = -\dfrac{j_c}{4F} M_{O_2} + \dfrac{j_c}{2F} M_{H_2O}$ |
| Momentum conservation equation | | $\nabla \cdot (\varepsilon \rho \vec{u} \vec{u}) = -\varepsilon \nabla P + \vec{\nabla} \cdot (\varepsilon \mu \nabla \vec{u}) + S_u$ | GDL and CL | $S_u = \dfrac{-\mu}{K_p} \varepsilon \vec{u}$ |
| Species transport equation | | $\nabla \cdot (\varepsilon \vec{u} C_i) = \nabla \cdot (D_i^{eff} \nabla C_i) + S_i$ | CL | $S_{H_2} = -\dfrac{j_a}{2F}$ <br> $S_{O_2} = -\dfrac{j_c}{4F}$ <br> $S_{H_2O} = +\dfrac{j_c}{2F}$ |
| Energy conservation equation | | $\nabla \cdot (\rho c_p \vec{u} T) = \nabla \cdot (k_{eff} \nabla T) + S_T$ | CL | $S_T = I^2 R_{ohm} + r_w h_l - \eta_{a.c} j_{a.c} + h_{reaction}$ |
| Butler–Volmer equation | Anode | $j_a = (\xi_a j_a^{ref})_a \left(\dfrac{C_{H2}}{C_{H2}^{ref}}\right)^{\gamma_a} \left[ \exp\left(\dfrac{\alpha_a F}{RT} \eta_a\right) - \exp\left(\dfrac{-\alpha_c F}{RT} \eta_a\right) \right]$ | - | - |
| | Cathode | $j_c = (\xi_c j_c^{ref})_c \left(\dfrac{C_{O2}}{C_{O2}^{ref}}\right)^{\gamma_c} \left[ -\exp\left(\dfrac{\alpha_a F}{RT} \eta_c\right) + \exp\left(\dfrac{-\alpha_c F}{RT} \eta_c\right) \right]$ | | |
| Current transport equation | Anode | $\nabla \cdot (\sigma_{sol} \cdot \nabla \phi_{sol}) - j_{sol} = 0$ <br> $\eta_a = \emptyset_{sol} - \emptyset_{mem}$ | Anode side | $j_s = -j_a < 0$ <br> $j_m = +j_a > 0$ |
| | Cathode | $\nabla \cdot (\sigma_{mem} \cdot \nabla \phi_{mem}) + j_{mem} = 0$ <br> $\eta_c = \emptyset_{sol} - \emptyset_{mem} - V_{OC}$ | Cathode side | $j_s = +j_c > 0$ <br> $j_m = -j_c < 0$ |
| Liquid water formation and transport | | $\nabla \cdot (\rho_l \vec{v_l} s) = r_w$ | Based on the operating condition | $r_w = c_r \max \begin{cases} (1-s) \dfrac{P_{wv} - P_{sat}}{RT} M_{w.H_2O} \\ -s \rho_l \end{cases}$ |

## 4- Theory

One way to determine the oxygen mass transport resistance is to use the limiting current density method [6,49,50]. When the cell operates in its limiting current, oxygen consumption is at the maximum rate. In other words, according to Equation 1 and Equation 2, the amount of oxygen transport resistance can be calculated by Equation 3. Figure 2 shows a schematic view of transport phenomena on the cathode side.

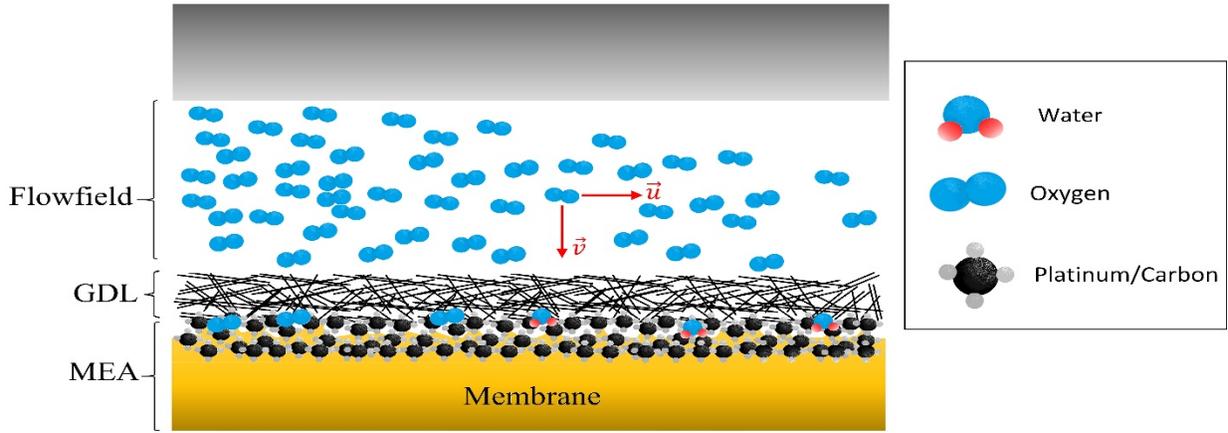

**Figure 2-** Schematic 2D view of transport phenomena at the cathode side

$$\dot{N}_{O_2.Transport} = \frac{C_{O_2}^{ch} - C_{O_2}^{pt}}{R_T} \tag{1}$$

$$\dot{N}_{O_2.Reaction} = \frac{i}{4F} \tag{2}$$

$$R_T = \frac{4 \times F \times C_{O_2}^{ch}}{i_{lim}} \tag{3}$$

The resistance in every component is determined according to the oxygen concentration. The normalized resistance is defined as follows:

$$\hat{R}_T = \frac{R_T}{R_0} \tag{4}$$

Where $R_0$ is the resistance of the case with 1 mol m$^{-3}$ inlet concentration of oxygen and limiting current density of 3.86 A cm$^{-2}$. The current density is selected as the maximum predicted limiting current density by simulations. The mass fluxes with convection $Q_{Convective}$ and diffusion $Q_{Diffusive}$ can be calculated by:

$$Q_{Convective} = \rho u \tag{5}$$

$$Q_{Diffusive} = -D \frac{\partial C}{\partial y} \tag{6}$$

The normalized quantities of mass fluxes are:

$$\hat{Q}_{Convective} = \frac{\rho u}{\rho u_0} \tag{7}$$

$$\hat{Q}_{Diffusive} = \frac{-D \frac{\partial C}{\partial y}}{-D_0 \frac{\partial C_0}{\partial y}} \tag{8}$$

$D_0 \frac{dC_0}{dy}$ and $u_0$ are assumed $1.1 \times 10^{-4}$ and $1 \times 10^{-3}$, respectively. Constant parameters are the highest amount calculated in simulations.

The present study takes into account the transport phenomena within the cathode side of the PEMFC precisely. The study introduces polynomial regressions, predicting the phenomena. Then, the dominant features, including the porosity of GDL, stoichiometries of both sides, operating pressure, and temperature, are optimized. The features are critical factors in controlling oxygen and water management [6,16,49].

## 5- Numerical procedure

In the present investigation, a single cell PEMFC is modelled with a combination of RSM and 3D multiphase steady-state CFD. In CFD, the transport equations are discretized by FVM. A SIMPLE algorithm is employed pressure-velocity coupling. RSM can offer a proper understanding in a short period. The advantage of RSM in developing fuel cells is that some responses such as power density can be related to multiple input parameters or factors. RSM consists of statistical techniques that have been applied successfully to design experiments, create models, and determine the effects of factors [5,51].

RSM is employed to produce second-order polynomial regression. The second-order polynomial equation can predict the effect of features interaction. The form of the equation is:

$$y = \beta_0 + \sum_{i=1}^{j} \beta_i x_i + \sum_{i<j}^{j}\sum \beta_{ji}x_ix_j + \sum_{i=1}^{j} \beta_{ii}x_{ii}^2 + \text{error} \tag{9}$$

Where y is the response, and the features are $x_i$. $\beta_0$ and $\beta_i$ ( $i \neq 0$ ) represent intercept and coefficient. First, a set of n cases must be run where independent factors are set to achieve y. Central Composite Design (CCD) defines the design matrix for RSM. The matrix-type of the model is expressed as:

$$\begin{bmatrix} y_1 \\ y_2 \\ \cdot \\ \cdot \\ \cdot \\ y_n \end{bmatrix} = \begin{bmatrix} 1 & x_{11} & x_{12} & \cdot & \cdot & \cdot & x_{1k} \\ 1 & x_{21} & x_{22} & \cdot & \cdot & \cdot & x_{2k} \\ \cdot & \cdot & & & & & \cdot \\ \cdot & \cdot & & & & & \cdot \\ \cdot & \cdot & & & & & \cdot \\ 1 & x_{n1} & x_{n2} & \cdot & \cdot & \cdot & x_{nk} \end{bmatrix} \begin{bmatrix} \beta_1 \\ \beta_2 \\ \cdot \\ \cdot \\ \cdot \\ \beta_k \end{bmatrix} + \begin{bmatrix} \varepsilon_1 \\ \varepsilon_2 \\ \cdot \\ \cdot \\ \cdot \\ \varepsilon_n \end{bmatrix} \tag{10}$$

The least-squares is applied to solve the equation. The error is assumed to be random, and this method's mean value is zero. Error $\varepsilon_i$ represents the difference between the observed ($y_i$) and calculated responses ($\hat{y}_i$).

$$\varepsilon_i = y_i - \hat{y}_i \tag{11}$$

In this case, the sum of squares of errors (SSE) is achieved as described below:

$$SSE = \sum_{i=1}^{n} \varepsilon_i^2 = \sum (y_i - \hat{y}_i)^2 \tag{12}$$

The error and SSE are stated as follows:

$$\varepsilon = y - x\beta \tag{13}$$

$$SSE = \varepsilon^T \varepsilon = (y - x\beta)^T (y - x\beta) \tag{14}$$

The objective is to select the β that minimizes the SSE in the equation. When the equations are derived based on β, a vector of partial derivatives are defined by the following equation:

$$\frac{\delta}{\delta \beta}(SSE) = -2x^T(y - x\beta) \tag{15}$$

If this derivative is zero, the expression xβ = y is found. The equation is solved to obtain the coefficients:

$$x^T x\beta = x^T y \tag{16}$$

$$\beta = (x^T x)^{-1} x^T y \tag{17}$$

The features are selected based on the studies which introduced the most significant features at low voltages [6,16,49]. The employed parameters are operating temperature, operating pressure, the stoichiometries of both sides, and the porosity of GDL. Central Composite Design (CCD) was utilized to set a design matrix. The levels of features are tabulated in Table 3, and Table 4 shows the design matrix. Objectives are normalized resistance and water saturation at GDL and flow field resistance. Lower resistance does not guarantee ideal water management, and in some cases, lowering resistance and ignoring water saturation can bring flooding. Therefore, to reach the best transport phenomena, these objectives are selected.

Table 3- Features used by RSM design

| Parameter | Unit | Level | | |
|---|---|---|---|---|
| | | low | Medium | High |
| $\varepsilon$ | ° | 0.1 | 0.5 | 0.9 |
| $S_c$ | - | 1 | 2 | 3 |
| $S_a$ | - | 1 | 2 | 3 |
| P | atm | 1 | 2.5 | 4 |
| T | K | 298 | 310.5 | 323 |

**Table 4-** Design matrix of RSM

| | T | Sa | Sc | P | Porosity |
|---|---|---|---|---|---|
| case 1 | 298 | 1 | 1 | 1 | 0.9 |
| case 2 | 298 | 3 | 1 | 1 | 0.1 |
| case 3 | 298 | 1 | 3 | 1 | 0.1 |
| case 4 | 298 | 3 | 3 | 1 | 0.9 |
| case 5 | 298 | 1 | 1 | 4 | 0.1 |
| case 6 | 298 | 3 | 1 | 4 | 0.9 |
| case 7 | 298 | 1 | 3 | 4 | 0.9 |
| case 8 | 298 | 3 | 3 | 4 | 0.1 |
| case 9 | 323 | 1 | 1 | 1 | 0.1 |
| case 10 | 323 | 3 | 1 | 1 | 0.9 |
| case 11 | 323 | 1 | 3 | 1 | 0.9 |
| case 12 | 323 | 3 | 3 | 1 | 0.1 |
| case 13 | 323 | 1 | 1 | 4 | 0.9 |
| case 14 | 323 | 3 | 1 | 4 | 0.1 |
| case 15 | 323 | 1 | 3 | 4 | 0.1 |
| case 16 | 323 | 3 | 3 | 4 | 0.9 |
| case 17 | 310.5 | 1 | 2 | 2.5 | 0.5 |
| case 18 | 310.5 | 3 | 2 | 2.5 | 0.5 |
| case 19 | 310.5 | 2 | 1 | 2.5 | 0.5 |
| case 20 | 310.5 | 2 | 3 | 2.5 | 0.5 |
| case 21 | 310.5 | 2 | 2 | 1 | 0.5 |
| case 22 | 310.5 | 2 | 2 | 4 | 0.5 |
| case 23 | 298 | 2 | 2 | 2.5 | 0.5 |
| case 24 | 323 | 2 | 2 | 2.5 | 0.5 |
| case 25 | 310.5 | 2 | 2 | 2.5 | 0.1 |
| case 26 | 310.5 | 2 | 2 | 2.5 | 0.9 |
| case 27 | 310.5 | 2 | 2 | 2.5 | 0.5 |

Two metrics are presented to evaluate the regression. R-squared ($R^2$) and Mean Square Error (MSE) are the metrics. The metrics are calculated by:

$$\text{MSE} = \frac{\sum_{i=1}^{N}(y_i - \hat{y}_i)^2}{n} \tag{18}$$

$$R^2 = 1 - \frac{\sum_{i=1}^{N}(y_i - \hat{y}_i)^2}{\sum_{i=1}^{N}(y_i - \bar{y})^2} \tag{19}$$

At the next step, the equation is utilized for prediction and MOO. MOO includes two layers. The first layer creates a Pareto front, and the second layer selects an ideal point from the front. MOPSO and NSGA-II are utilized to produce two different fronts which can challenge each other, and TOPSIS is the selection technique in the second layer. The comprehensive view of the investigation is presented in Figure 3 (a). The algorithms of NSGA-II and TOPSIS are presented in our previous study [5]. The iteration algorithm of MOPSO is shown in Figure 3 (b).

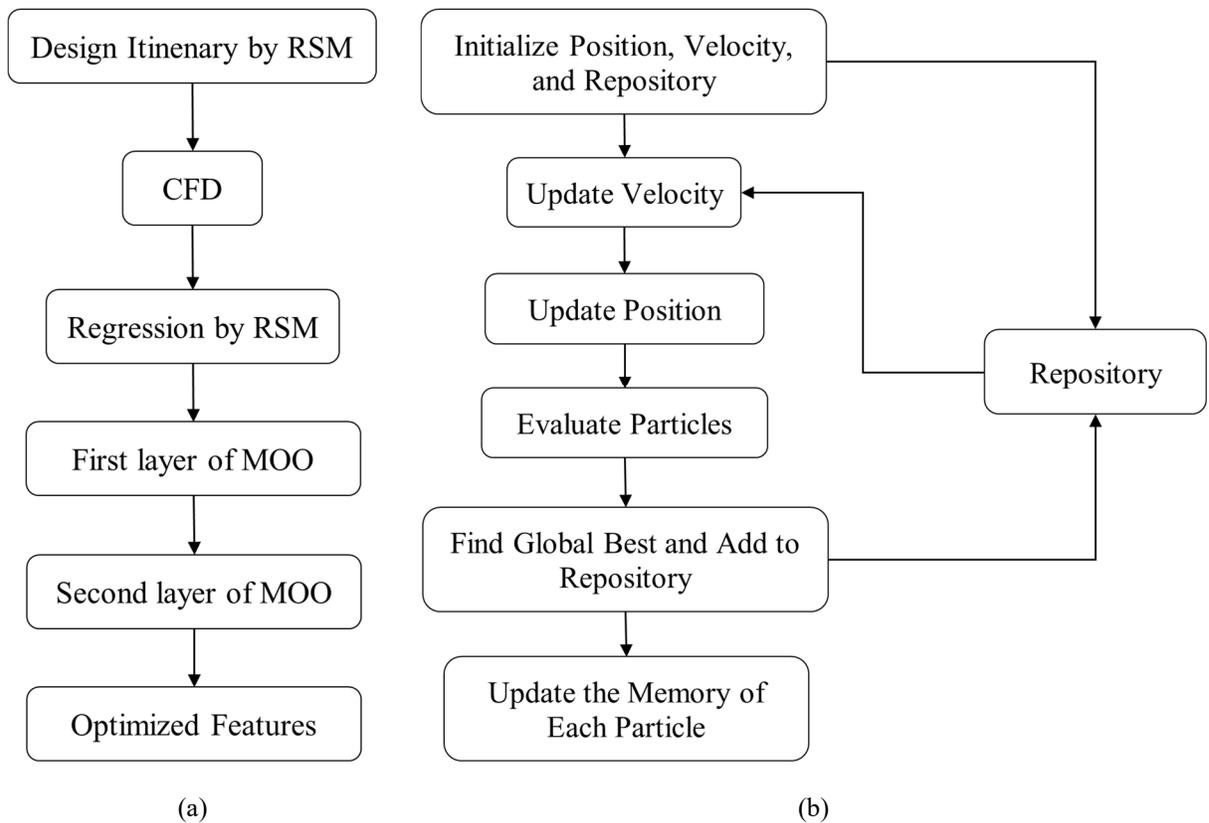

**Figure 3-** Flowchart of the present study (a) comprehensive view (b) MOPSO iteration algorithm

## 6- Grid independency and validation

The computational grid in the simulation has hexagonal elements reducing the distortion. Figure 4 (a) shows the mesh. The grid independency was checked by increasing the number of nodes. Different computational grids with various element counts were analyzed, including 4010651, 1944360, 710193, 350111, and 180324 were analyzed. Figure 4 (b) shows that the sample simulation with 1944360 elements was selected as the appropriate grid.

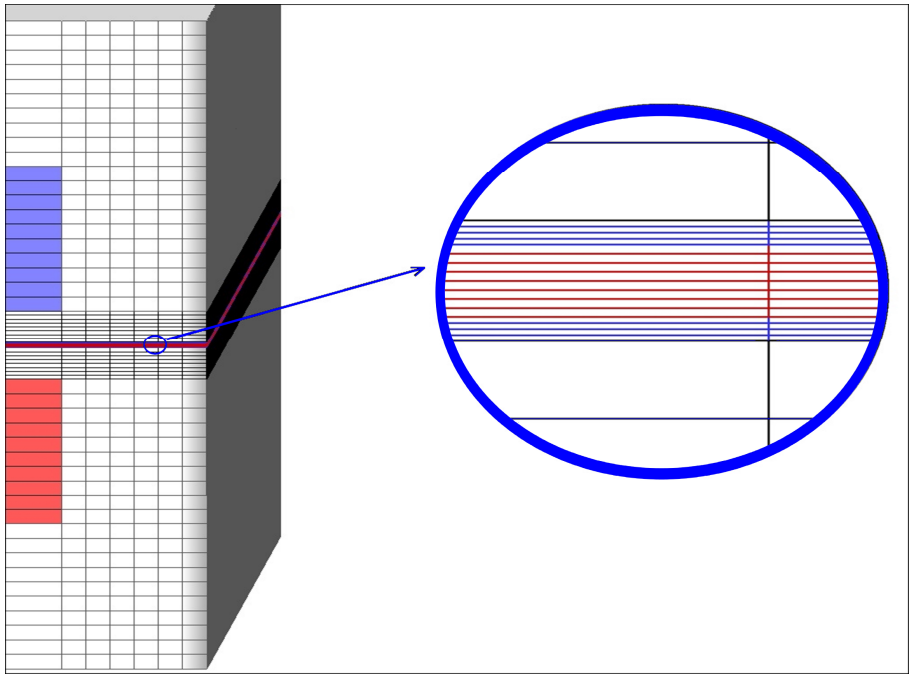

(a)

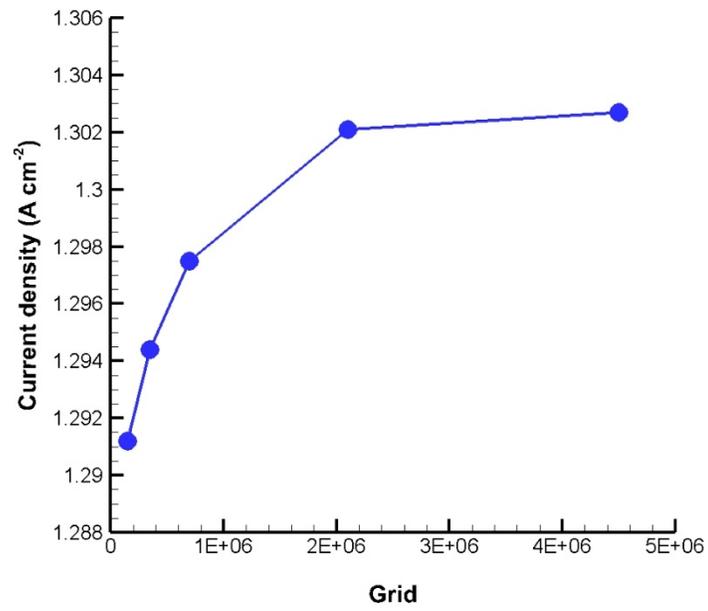

(b)

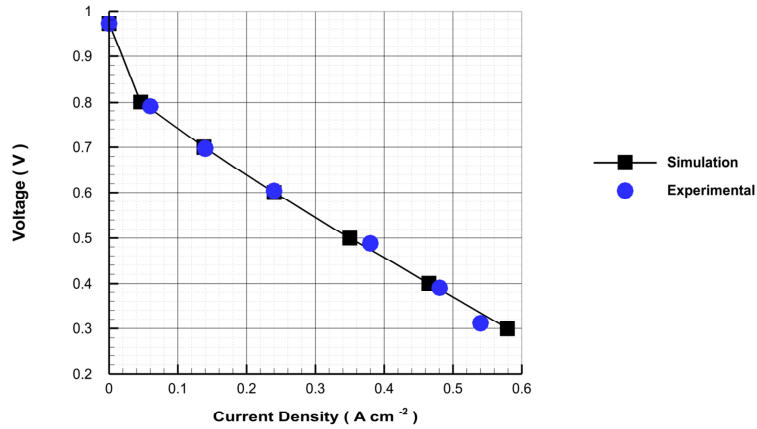

(c)

Figure 4- Simulation (a) structured mesh (b) grid independency and (c) validation

To evaluate the credibility of the model, the findings of the simulation are compared with experimental data presented in reference [52]. The geometry, operational conditions, and physical properties of materials are identical to the experiment. Figure 4 (c) shows appropriate coherence between the simulation and the reference [52].

## 7- Results

Two regressions for the resistance and water saturation are presented, and their accuracy is studied. Objectives are investigated in different conditions and minimized.

### 7-1- Case study

This section compares a simple parallel flow field with a tapered parallel flow field. Simulation parameters are 323 K, 4 atm, GDL porosity 0.5, and both sides stoichiometries 3. The parameters were resulted from taking both RSM levels and reference [5]. The difference between the two flow fields is the inclination of the main channels. Ratio $\frac{W_{start}}{W_{end}}$ equals zero and 4 for simple and tapered flow fields, respectively. The inlet and outlet are adjusted. Figure 5 shows tapering enhanced the uniformity of velocity distribution between 21 channels resulting in a more uniform water saturation distribution at the reaction site. The uniformity increases limiting current density by 41 % from 0.88 A cm$^{-2}$ to 1.24 A cm$^{-2}$. The tapered parallel flow field is selected for the optimization.

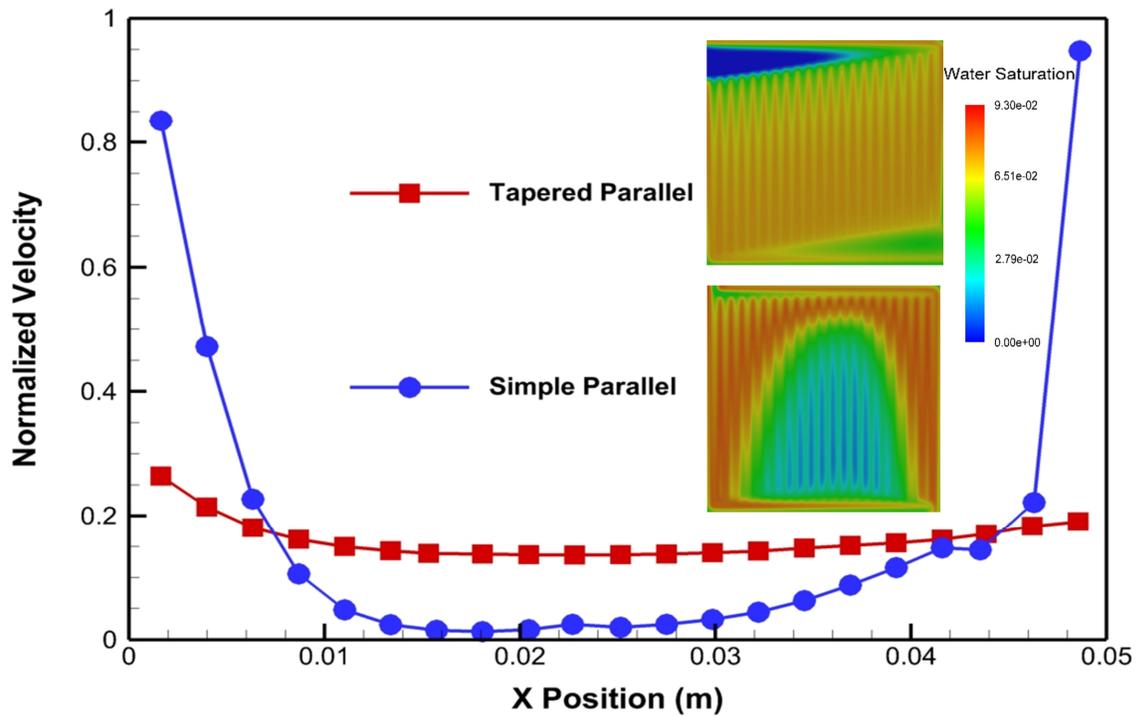

**Figure 5-** The effect of the flow field

**7-2- Polynomial Regression**

The optimization objectives are the normalized total resistance and water saturation at the interface of GDL and the flow field. Two polynomial equations are presented for the objectives. The proposed equations can be employed to study the interaction between electrochemical and dynamics of produced water in a reliable, straightforward way. The equations are:

$$S_{flowfield-GDL} = 1 \times 10^{-7} \, ( 11658321 - 33973 \, T + 1480205 \, Sa +$$
$$691175 \, Sc - 2702208 \, P - 10736769 \, Prosity - 7 \, T \times T - 16259 \, Sa \times$$
$$Sa + 9575 \, Sc \times Sc - 59447 \, P \times P + 1277181 \, Prosity \times Prosity -$$
$$3381 \, T \times Sa - 1515 \, T \times Sc + 9804 \, T \times P + 29858 \, T \times Prosity -$$
$$78703 \, Sa \times Sc - 50219 \, Sa \times P - 220686 \, Sa \times Prosity - 19480 \, Sc \times$$
$$P - 192390 \, Sc \times Prosity + 241073 \, P \times Prosity \,)$$

(20)

$$\hat{R}_T = 1 \times 10^{-7} \, ( 6389556345 - 37057983 \, T - 195204554 \, Sa -$$
$$221584443 \, Sc + 201480453 \, P - 768824004 \, Prosity + 54898 \, T \times T +$$
$$4945753 \, Sa \times Sa + 22376303 \, Sc \times Sc + 184888 \, P \times P +$$
$$105635187 \, Prosity \times Prosity + 500673 \, T \times Sa + 330637 \, T \times Sc -$$
$$212549 \, T \times P + 1910895 \, T \times Prosity + 5861312 \, Sa \times Sc +$$
$$4040490 \, Sa \times P + 7552381 \, Sa \times Prosity - 7893300 \, Sc \times P +$$
$$26388538 \, Sc \times Prosity - 20308980 \, P \times Prosity \,)$$

(21)

Two evaluating parameters, R-squared and MSE, are shown in Table 5. The errors of responses are presented in Figure 6. The results show that the regressions are reliable because of low and independent errors.

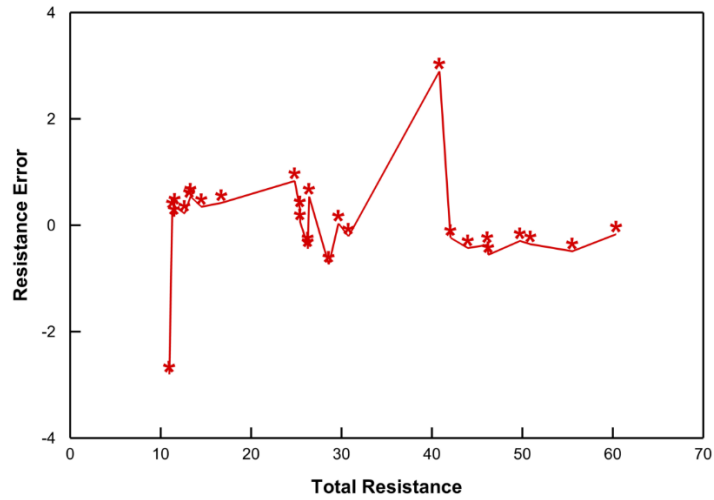

(a)

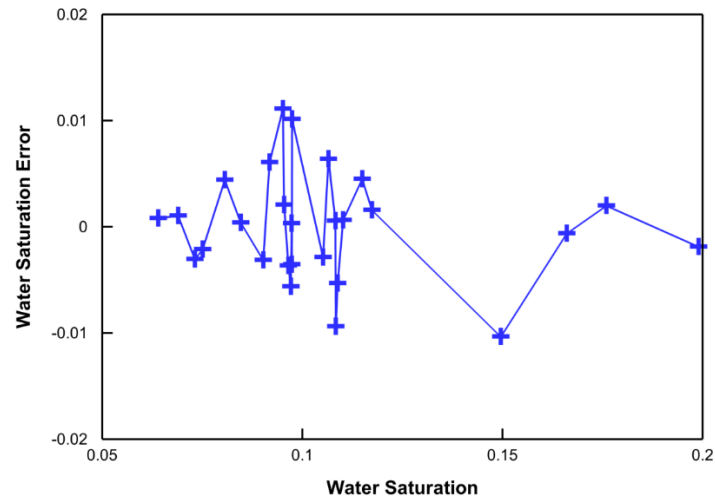

(b)

**Figure 6-** The residual plot of (a) Resistance (b) Water saturation

**Table 5-** Metrics to evaluate regression models

|  | **Normalized Resistance** | **Water Saturation** |
| --- | --- | --- |
| **$R^2$** | 0.9967 | 0.974 |

|     | MSE |     | 0.76 |     | $2.5 \times 10^{-5}$ |
| --- | --- | --- | --- | --- | --- |

The analysis of variances was carried out through RSM. In the analysis of variances, the sum of the squared deviation of each parameter and error is calculated first, followed by the degree of freedom, mean square, F-value, and P-value. All of the parameters are significant because the parameter itself or its interaction with the other parameters has a suitable high F-value and low P-value. The amount of F-value and P-value are tabulated in Table 6. DOF represents the degree of freedom. The most decisive parameter for water saturation and resistance are porosity and pressure. Interaction between parameters is of significance for water saturation in some interactions. If the interactions are illustrated by the signs of parameters, the decisive interactions include $\varepsilon - P$, $\varepsilon - T$, $\varepsilon - S_a$, $\varepsilon - S_c$, $S_c - S_a$, $S_a - P$, and $P - T$. The most significant interaction is $P - T$.

Table 6- The analysis of variances

|  | Resistance | | | Water saturation | | |
| --- | --- | --- | --- | --- | --- | --- |
|  | DOF | F value | P value | DOF | F value | P value |
| T | 1 | 6.36 | 0.045 | 1 | 18.33 | 0.005 |
| Sa | 1 | 1.67 | 0.243 | 1 | 1.26 | 0.304 |
| Sc | 1 | 30.90 | 0.001 | 1 | 2.97 | 0.135 |
| P | 1 | 1664.02 | 0.000 | 1 | 2.43 | 0.17 |
| Prosity | 1 | 23.43 | 0.003 | 1 | 43.53 | 0.001 |
| T×T | 1 | 1.83 | 0.496 | 1 | 0 | 0.988 |
| Sa×Sa | 1 | 0.80 | 0.690 | 1 | 0.06 | 0.818 |
| Sc×Sc | 1 | 0.74 | 0.107 | 1 | 0.02 | 0.892 |
| P×P | 1 | 4.27 | 0.973 | 1 | 3.89 | 0.096 |
| Prosity×Prosity | 1 | 1.61 | 0.203 | 1 | 9.08 | 0.024 |
| T×Sa | 1 | 1.72 | 0.225 | 1 | 2.54 | 0.162 |
| T×Sc | 1 | 0.43 | 0.406 | 1 | 0.51 | 0.502 |
| T×P | 1 | 6.56 | 0.422 | 1 | 48.12 | 0 |
| T×Prosity | 1 | 5.21 | 0.084 | 1 | 31.73 | 0.001 |
| Sa×Sc | 1 | 6.95 | 0.252 | 1 | 8.82 | 0.025 |
| Sa×P | 1 | 6.36 | 0.238 | 1 | 8.08 | 0.029 |
| Sa×Prosity | 1 | 1.67 | 0.538 | 1 | 11.09 | 0.016 |
| Sc×P | 1 | 30.90 | 0.043 | 1 | 1.22 | 0.312 |
| Sc×Prosity | 1 | 1664.02 | 0.063 | 1 | 8.43 | 0.027 |
| P×Prosity | 1 | 23.43 | 0.039 | 1 | 29.79 | 0.002 |

### 7-3- The prediction of effects

The influence of features on the objectives, current density, water management, and mass transport phenomena are investigated. The condition of the reaction site, output current density, and mass fluxes can explain the changes in the resistance and water saturation at the interface of GDL and the flow field. The conditions of the reaction site affect MEA resistance and the rate of oxygen consumption. The mass flux within GDL changes GDL resistance. The fixed amount features are porosity 0.68, temperature 323 K, pressure 1 atm, cathode stoichiometry 3, and anode stoichiometry 2.62. In the last stage of this paper, it will be determined that these values are optimum. For the resistance, interactions are not as important as the solitary effect of parameters. Significant interactions for water saturation are the combination $\varepsilon - P$, $\varepsilon - T$, $\varepsilon - S_a$, $\varepsilon - S_c$, $S_c - S_a$, $S_a - P$, and P- T, and they will be shown. The selection of axis for parameters is adjusted to bring a better view.

### 7-3-1- Pressure

Pressure is the most decisive parameter for the resistance. The effect of operating pressure on the transport phenomena is presented in Figure 7. The figure shows that water saturation is lower in low pressure. Furthermore, the normalized resistance in all components increases by the rise of pressure, and GDL has the lowest amount of resistance. The amendment of mass flux in different pressures is a decisive feature in the resistance in GDL. The oxygen concentration gradient increases through GDL thickness as pressure increases, but the diffusivity coefficient decreases. It leads to a maximum point on the normalized diffusive mass flux curve. On the other hand, the amount of normalized convective mass flux drops by pressure increment. As a result, the resistance in GDL increases.

The condition of the reaction site at different pressure determines a vital part of transport phenomena. At high pressure, the amount of water saturation is higher, the uniformity of oxygen at the reaction site is low, and the concentration of unconsumed oxygen at the reaction site is more. Although, after medium pressure, the amount of current density decreases, the amount of oxygen at the reaction site increases, which is the sign of lower performance of MEA. Lower oxygen gradient and oxygen uniformity at reaction site, higher saturated water, and unconsumed oxygen caused higher resistance and worse water management.

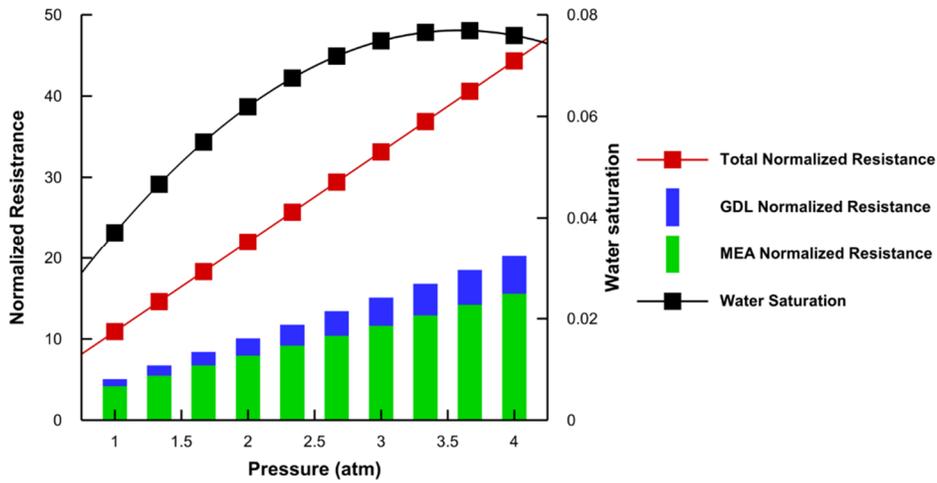

(a)

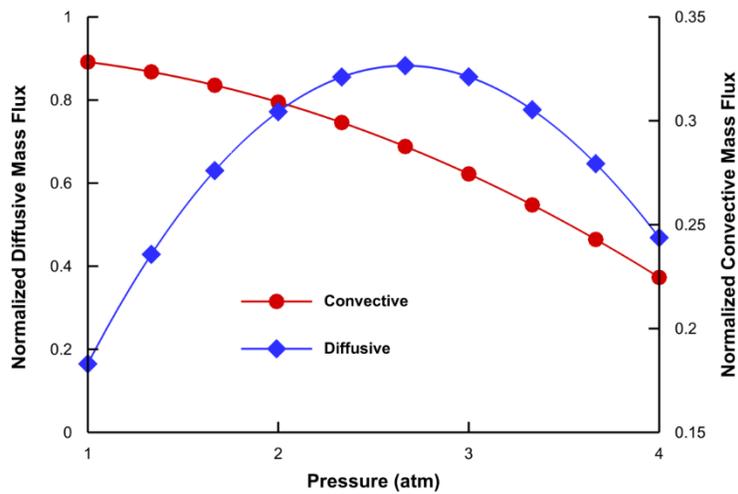

(b)

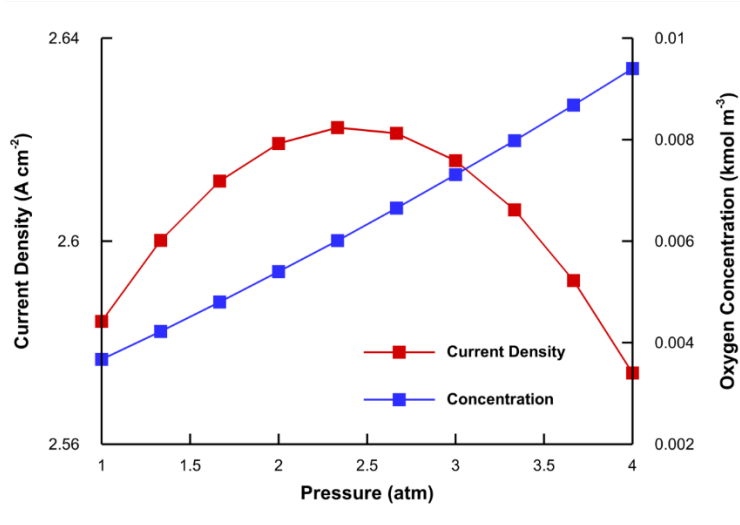

(c)

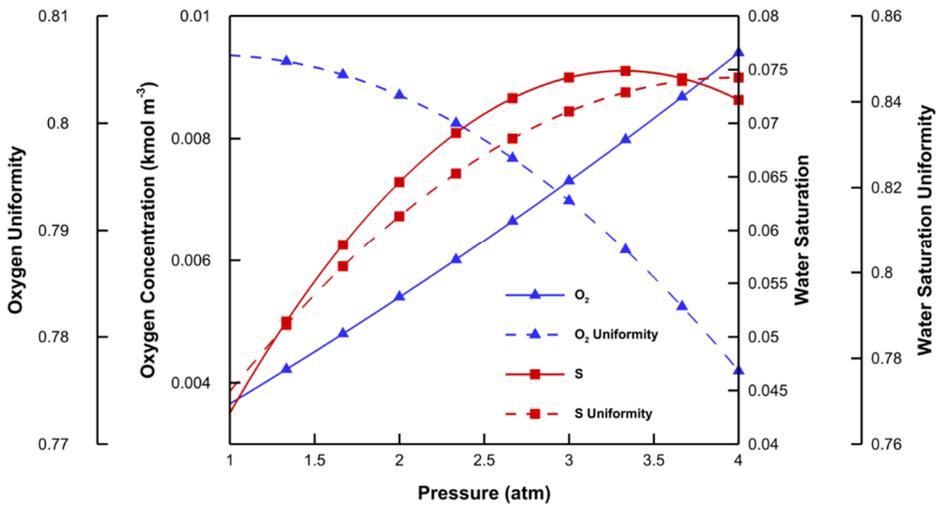

(d)

**Figure 7-** The effect of pressure on (a) normalized resistance and water saturation (b) convective and diffusive mass flux (c) current density and oxygen concentration at the reaction site (d) the distribution of reactants at the reaction site

Pressure affects water management, reaction rate, and mass flux. Figure 8 demonstrates the effect of essential interactions, in which pressure is one side of them. The interaction of pressure and porosity is the most significant interaction for the resistance. The interaction is almost one-way. Higher pressure increases the normalized resistance. Therefore, interactions are not so crucial for the resistance. The effect of pressure and temperature interaction on water saturation is essential. Pressure and temperature can change the rate of condensation, mass flux, reaction rate, and the amount of produced water. Lower pressure leads to lower water saturation at a high temperature, and at low temperature, the trends reverse. Minimum water saturation is 0.03 occurring at 323 K and 1 atm. The interaction of pressure and anode stoichiometry can change the cell's electrochemical behaviour, resulting in different water production and mass flux. Pressure 4 atm and anode stoichiometry 1 brings minimum water saturation.

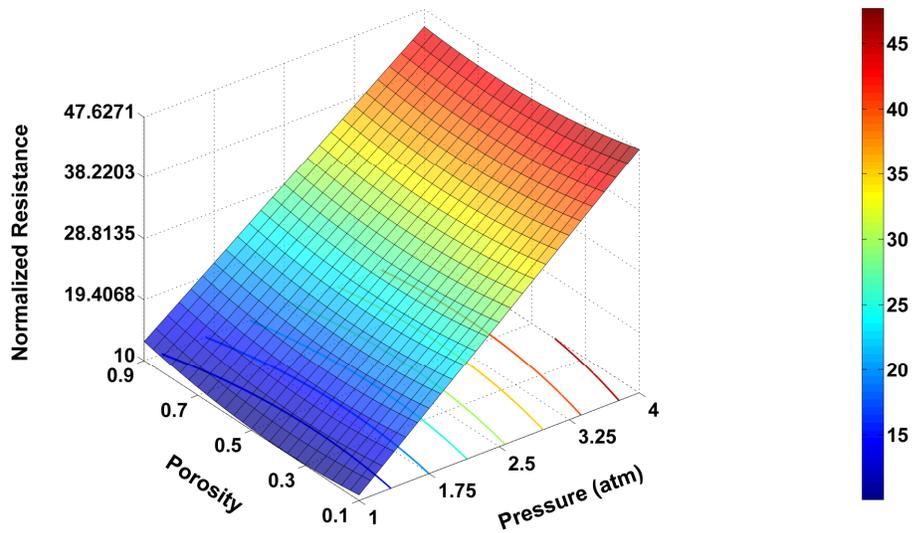

(a)

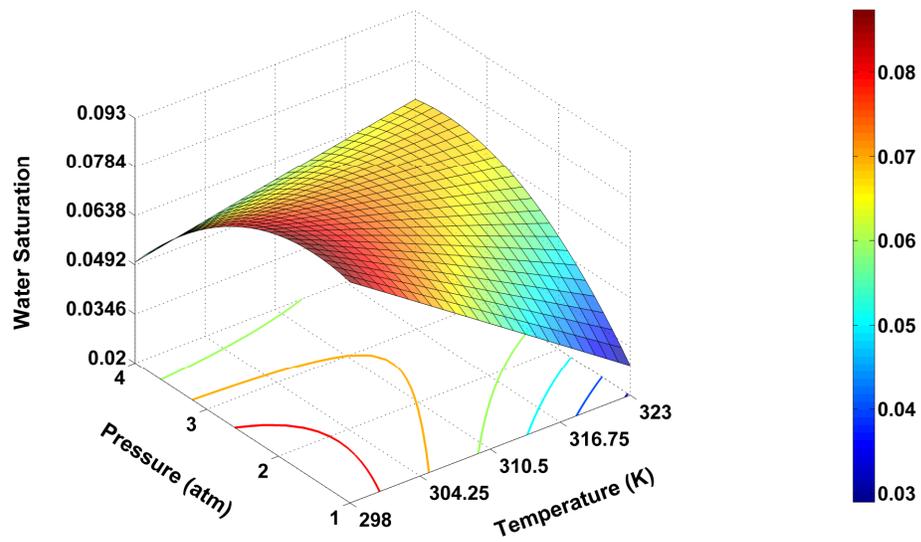

(b)

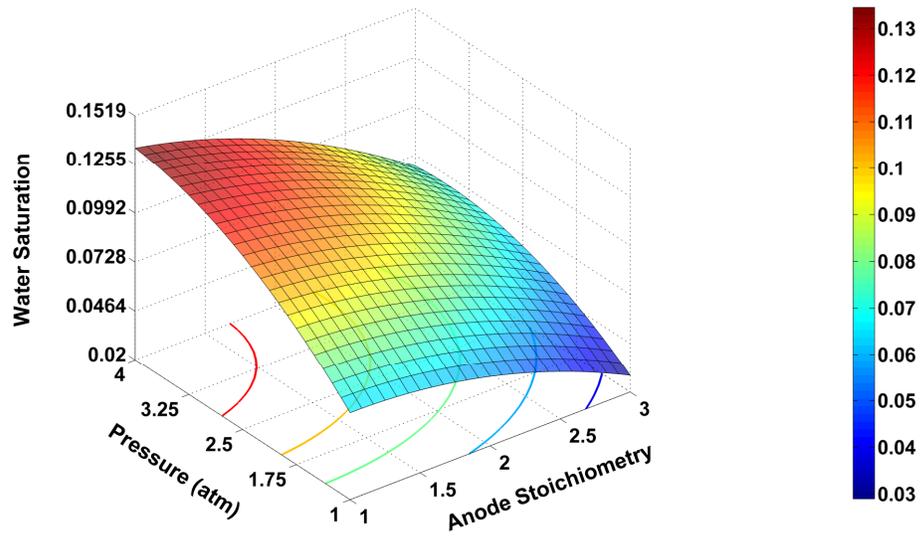

(c)

**Figure 8-** The effect of the interaction of pressure and (a) porosity on the normalized resistance (b) temperature on water saturation (c) anode stoichiometry on water saturation

### 7-3-4- Porosity

The porosity of GDL is the most critical parameter determining water saturation. The impact of porosity on the transport phenomena is presented in Figure 9. At low porosity, water saturation at the interface of the flow field and GDL is more. The lower amount of the normalized resistance occurs at porosity 0.37.

The change of GDL porosity can alter mass fluxes in GDL, consequently, the resistance and water management in GDL. Different porosities may change permeability in porous media. Higher porosity decreases the convective mass flow rate. There is a minimum point in medium porosity as an interaction of changes in diffusion coefficient and gradient of oxygen concentration in the direction perpendicular to the reaction site. As a result, the resistance of GDL drops by the increase of porosity.

Different porosity leads to different conditions in the reaction site. The current density changes in different porosity are identical to the oxygen concentration trend at the reaction site. The maximum amount of them is approximately at porosity 0.6. Higher porosity changes water saturation in a converse trend, in which the minimum amount of water saturation is at porosity 0.72. Low porosity brings higher water saturation with high uniformity and low unconsumed oxygen with uniformity at the reaction site. The cumulative result is that lower porosity leads to lower MEA resistance.

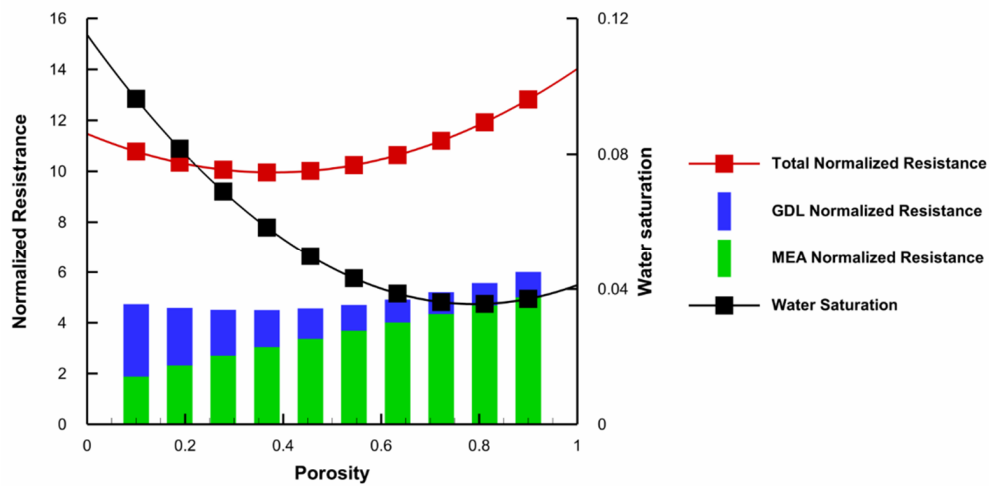

(a)

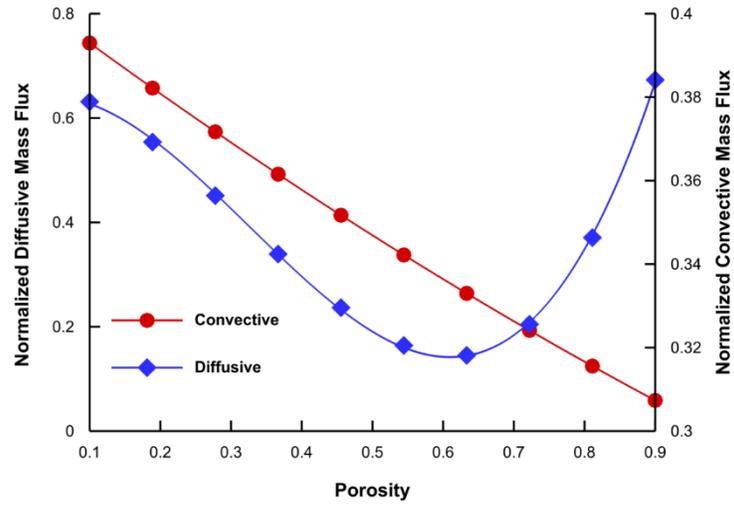

(b)

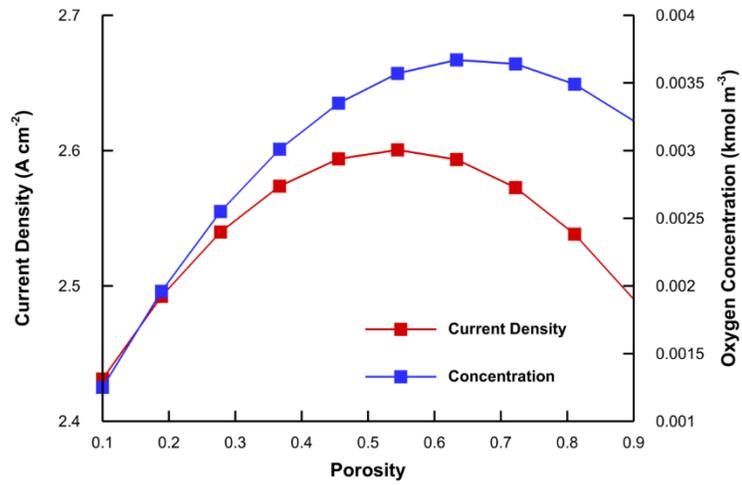

(c)

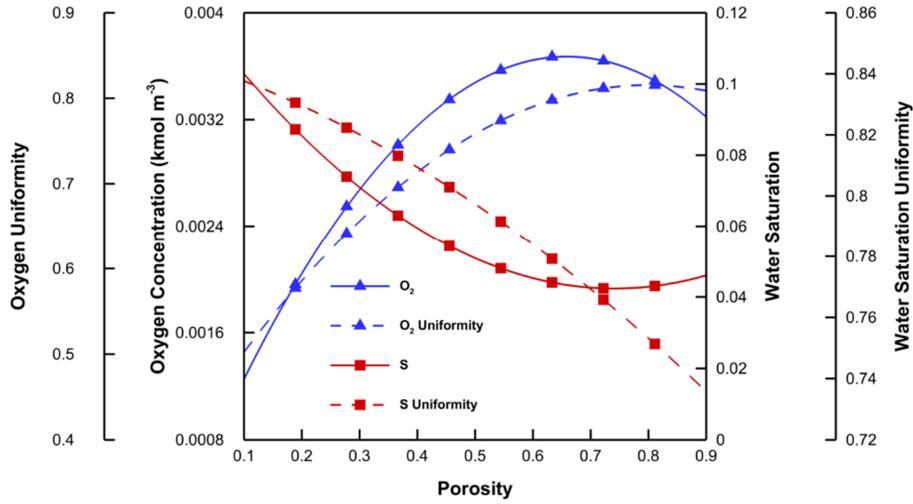

(d)

**Figure 9**- The effect of porosity on (a) normalized resistance and water saturation (b) convective and diffusive mass flux (c) current density and oxygen concentration at the reaction site (d) the distribution of reactants at the reaction site

The interaction of porosity and other parameters can facilitate or deteriorate the effect of porosity on the reaction site conditions. Figure 10 depicts the result of interactions. Medium porosity brings lower water saturation. Furthermore, medium porosity leads to the lowest water saturation at low pressure, high anode stoichiometry, low cathode stoichiometry, and high temperature.

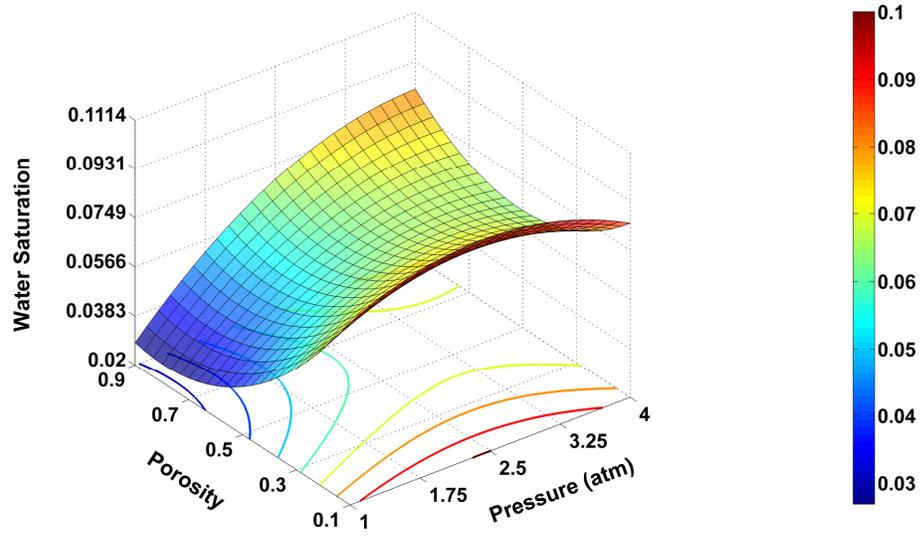

(a)

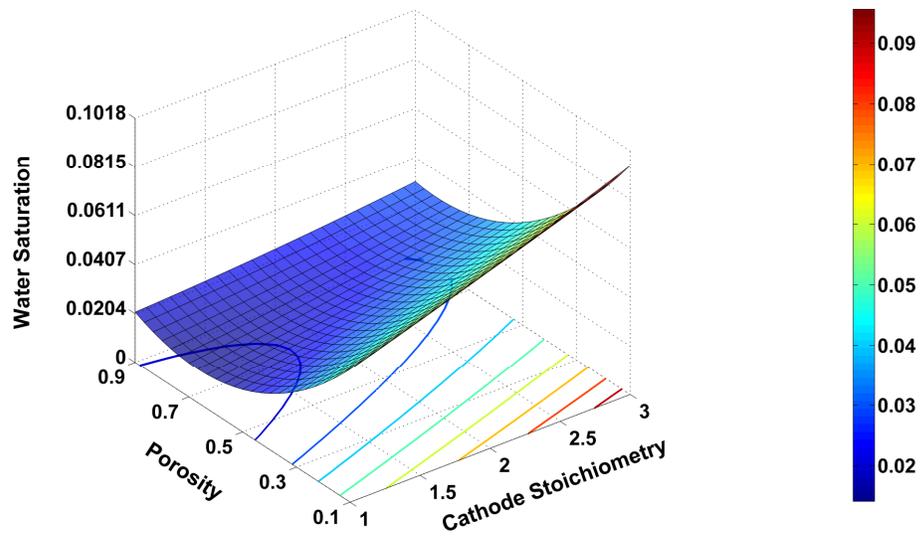

(b)

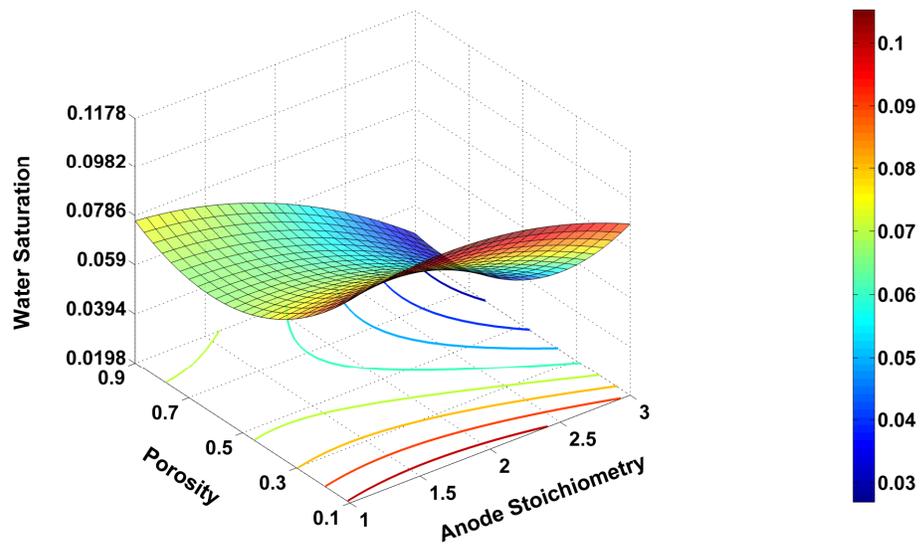

(c)

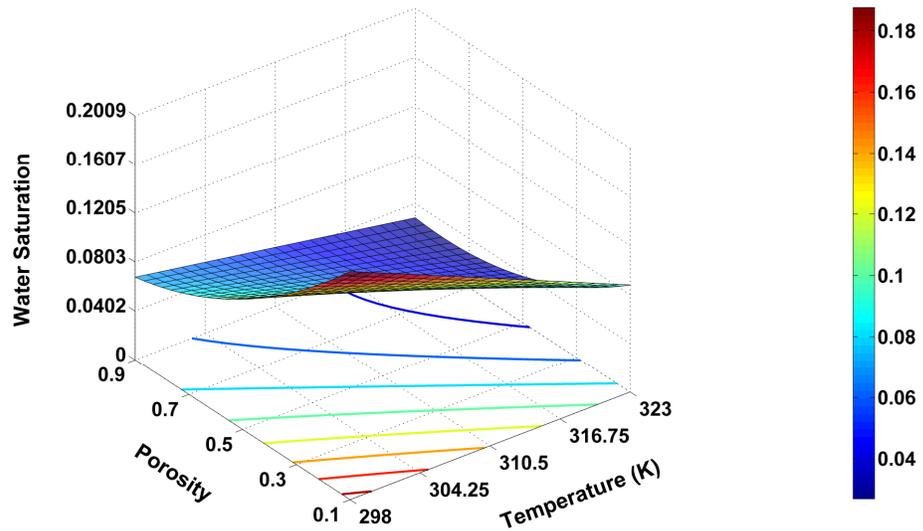

(d)

**Figure 10-** The effect of significant interactions, which one side of them is porosity on water saturation (a) porosity and pressure (b) porosity and cathode stoichiometry (c) porosity and anode stoichiometry (d) porosity and temperature

### 7-3-3- Cathode stoichiometry

A change in cathode stoichiometry changes the reaction site. The influence of cathode stoichiometry is shown in Figure 11. As stoichiometry increases, flow field resistance decreases, and MEA resistance increases. Interaction of changes causes a minimum amount of normalized resistance at medium cathode stoichiometry. Furthermore, the amount of water saturation on the interface of the flow field and GDL is higher in low cathode stoichiometry.

Cathode stoichiometry can change the performance of PEMFC. The current density has a maximum point with the change of cathode stoichiometry. At high cathode spirometry, the concentration of unconsumed oxygen is higher. Moreover, higher stoichiometry increases water removal. Overall, increasing cathode stoichiometry reduces water saturation at all interfaces while increasing MEA resistance.

The oxygen and water saturation condition is shown in Figure 11 (d). The uniformity of water saturation and oxygen distribution changes are directly related to the amount of water saturation and oxygen, respectively. Higher oxygen concentration results in higher oxygen uniformity, and lower water saturation leads to lower water saturation uniformity.

The effect of cathode stoichiometry on the mass flux condition is shown in Figure 11 (b). Normalized convective mass flux decreases. A higher inlet mass flow rate at the cathode side does not lead to higher convection. Furthermore, convective mass flux reaches a maximum at the medium cathode stoichiometry, and GDL resistance reaches its maximum point at cathode stoichiometry 2.3.

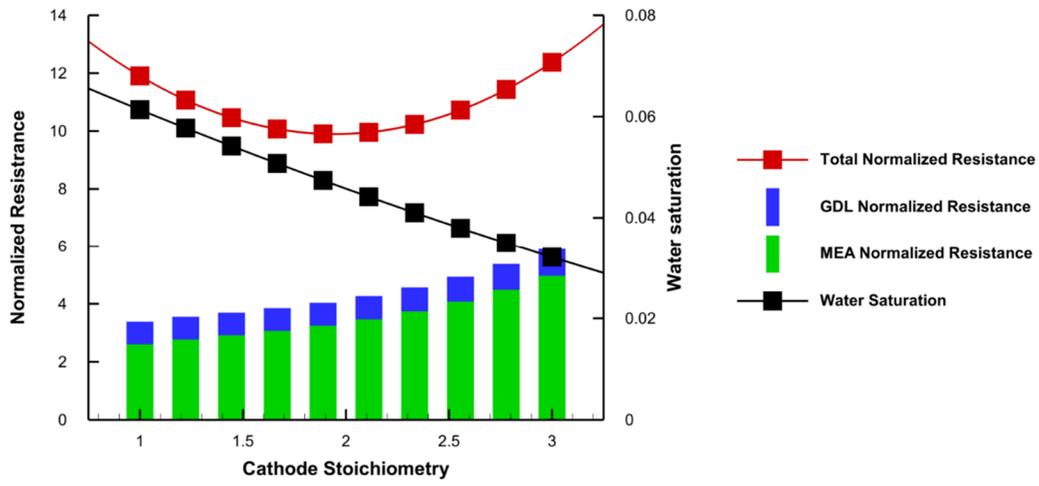

(a)

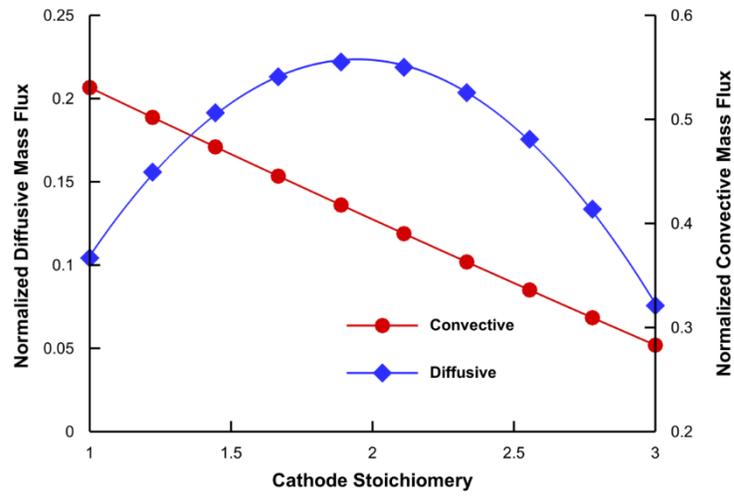

(b)

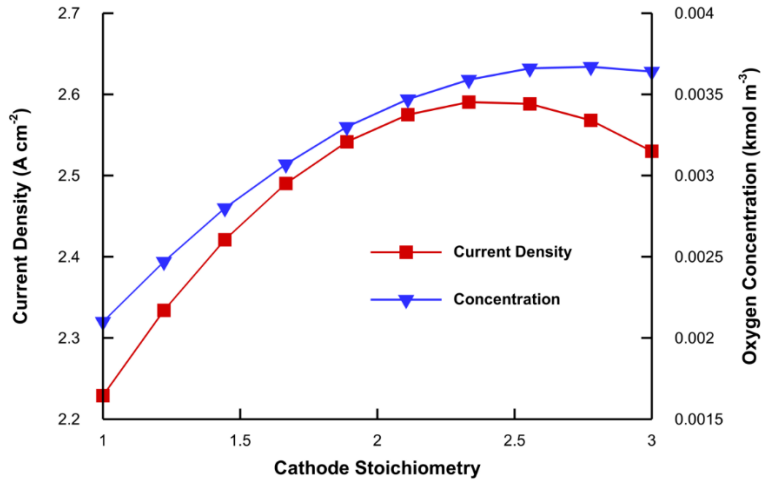

(c)

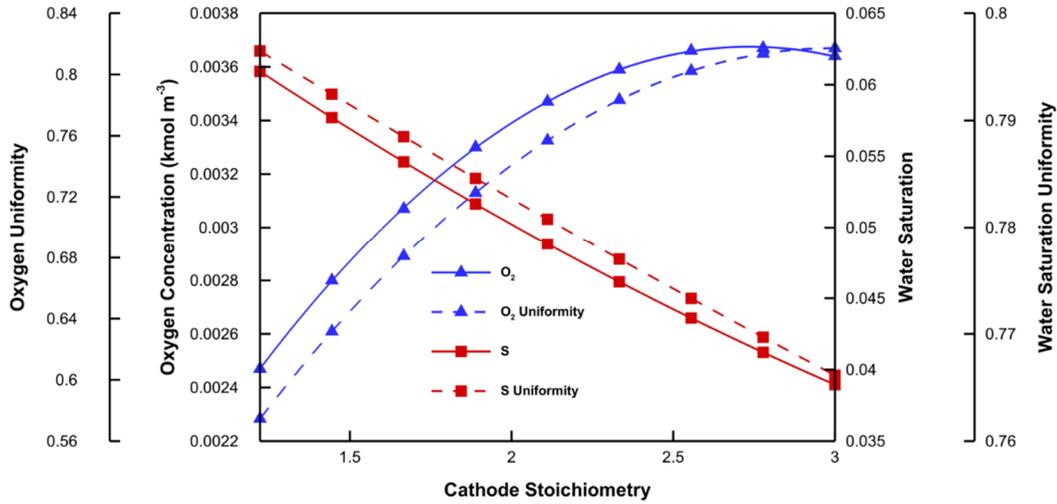

(d)

**Figure 11**- The effect of cathode stoichiometry on (a) normalized resistance and water saturation (b) convective and diffusive mass flux (c) current density and oxygen concentration at the reaction site (d) the distribution of reactants at the reaction site

The interaction of anode stoichiometry and cathode stoichiometry is decisive in determining water saturation. Figure 12 shows the effect of stoichiometries interaction on water saturation. Simultaneous high stoichiometries bring worse water management, and low stoichiometries lead to lower water saturation.

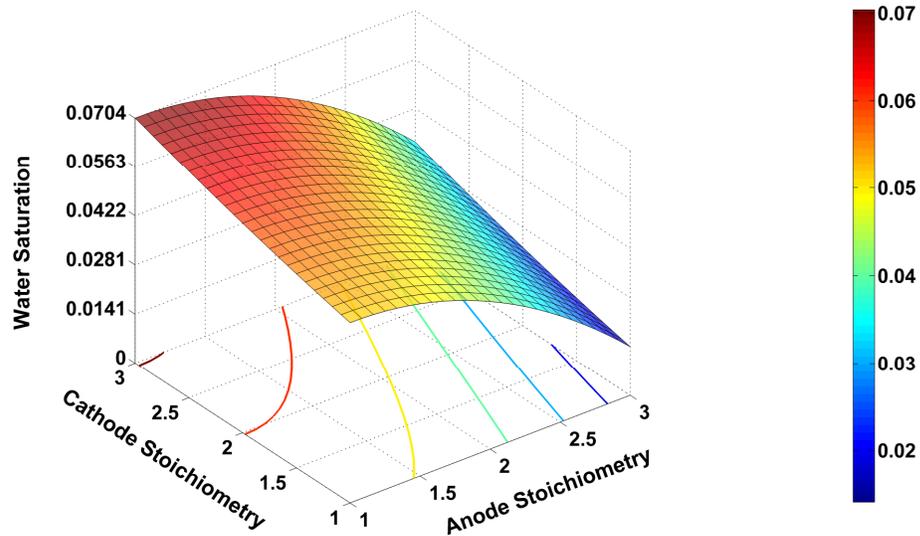

**Figure 12**- The interaction of cathode and anode stoichiometries

### 7-3-4- Anode stoichiometry

Higher anode stoichiometry can change the conditions of transport phenomena and water management in PEMFC, as shown in Figure 13. By increasing anode stoichiometry, water saturation decreases, and total normalized resistance increases. Furthermore, resistance in the flow field decreases.

The changes in mass flux are presented in Figure 13 (b). As anode stoichiometry grows, normalized convective mass flux rises, and normalized diffusive mass flux decreases. GDL resistance decreases in a trend similar to normalized convective mass flux.

The reaction site condition varies with anode stoichiometries. The change is in a trend that resembles cathode stoichiometry. With the increase of anode stoichiometry, the need for oxygen increases [5]. By increasing both stoichiometries, unconsumed oxygen increases, water saturation reduces, and MEA resistance increases. At anode stoichiometry 1.9, current density has its maximum amount. The uniformity of oxygen distribution is a function of two parameters, i.e. produced water saturation and oxygen consumption. The parameters are affected by anode stoichiometry because of the electrochemical behaviour of PEMFC. The interaction between these factors can cause an extremum point in the amount of uniformity.

At anode stoichiometry 1.65, the uniformity of oxygen concentration reaches a minimum and continues to increase after this turning point. When water saturation reduces, water saturation uniformity decreases.

The stoichiometry of the anode side affects the mass flux on the cathode side, and it is because of the change in the need for oxygen at the reaction site. The higher anode stoichiometry increases the convective mass flux and decreases diffusive mass flux. In a similar trend to diffusive mass flux, GDL resistance decreases.

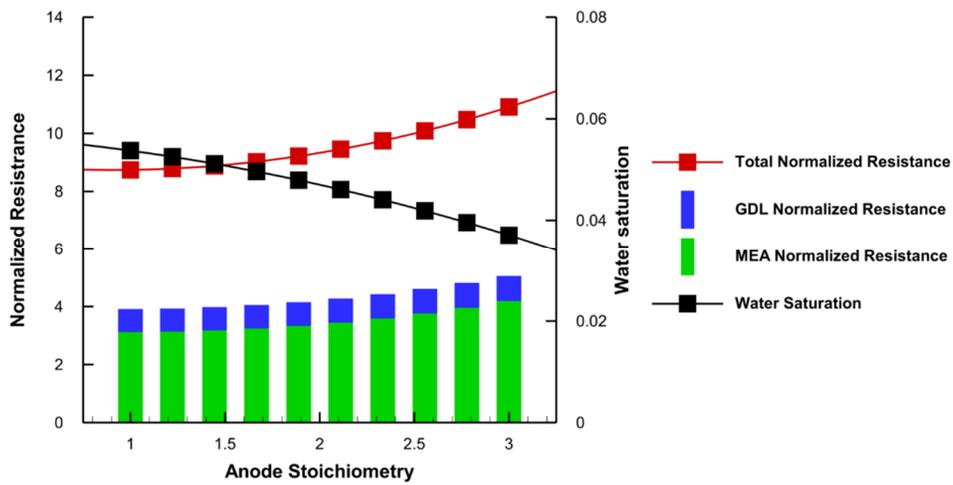

(a)

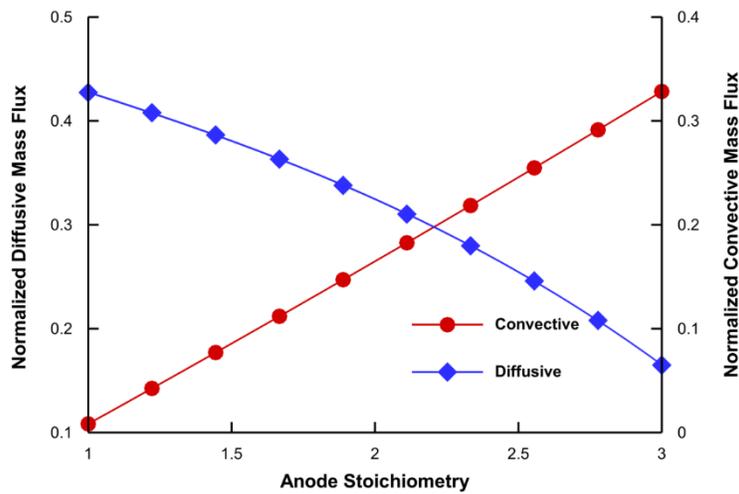

(b)

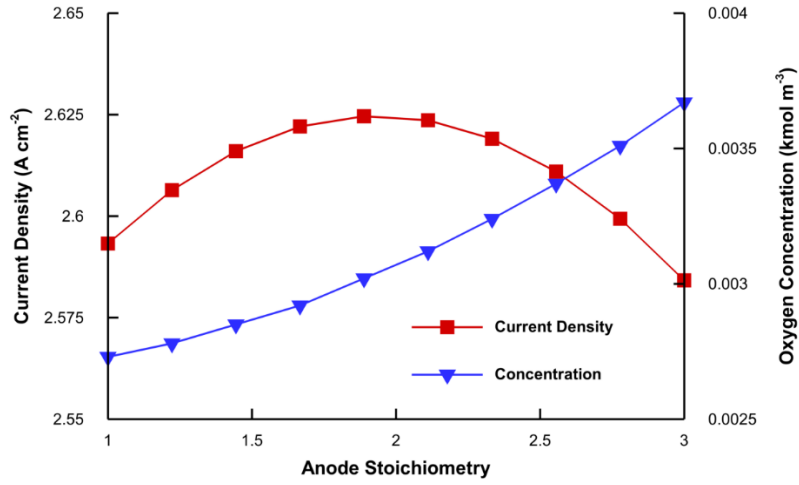

(c)

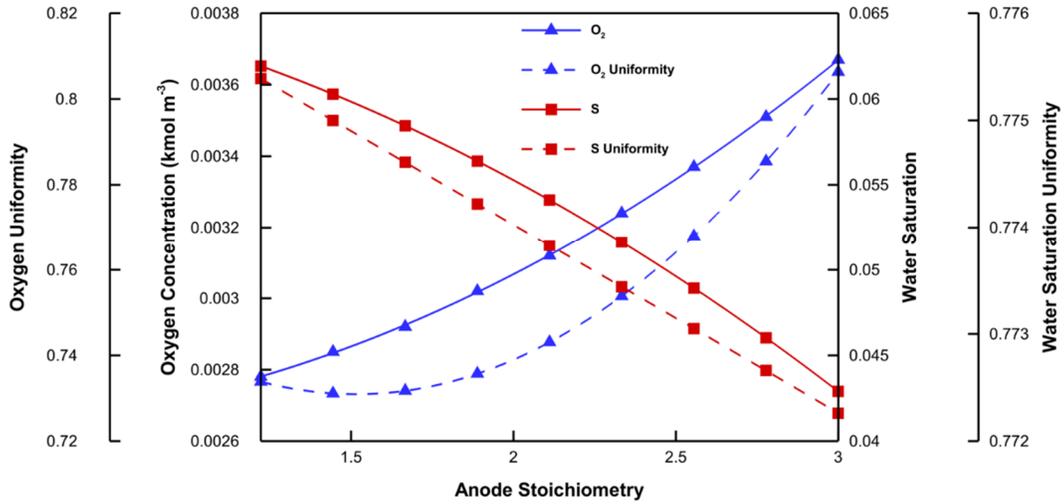

(d)

**Figure 13-** The effect of anode stoichiometry on (a) normalized resistance and water saturation (b) convective and diffusive mass flux (c) current density and oxygen concentration at the reaction site (d) the distribution of reactants at the reaction site

## 7-3-5- Temperature

Temperature can change transport phenomena in PEMFC. The influence of operating temperature on the transport phenomena is presented in Figure 14. With the increase in temperature, the water saturation decreases. The change of normalized resistance consists of two parts. First, from 298 K to 306 K, it reduces, and then from 306 K to 323 K, it rises. Flow field resistance is lower at a low level of temperature.

The mass fluxes in GDL can be changed by temperature. Temperature increases convective mass flux and decreases diffusive mass flux. In a similar trend to diffusive mass flux, GDL resistance drops.

Temperature affects the conditions in the reaction site. Temperature can change the probability of condensation and reaction rate [53]. At a temperature of 307 K, the current density has a maximum. However, as the need for oxygen decreases the unconsumed oxygen concentration at the reaction site increases. By increasing oxygen, the uniformity of oxygen increases. Higher temperature results in lower water saturation and lower uniformity of water saturation. At a temperature of 306 K, MEA resistance has a minimum.

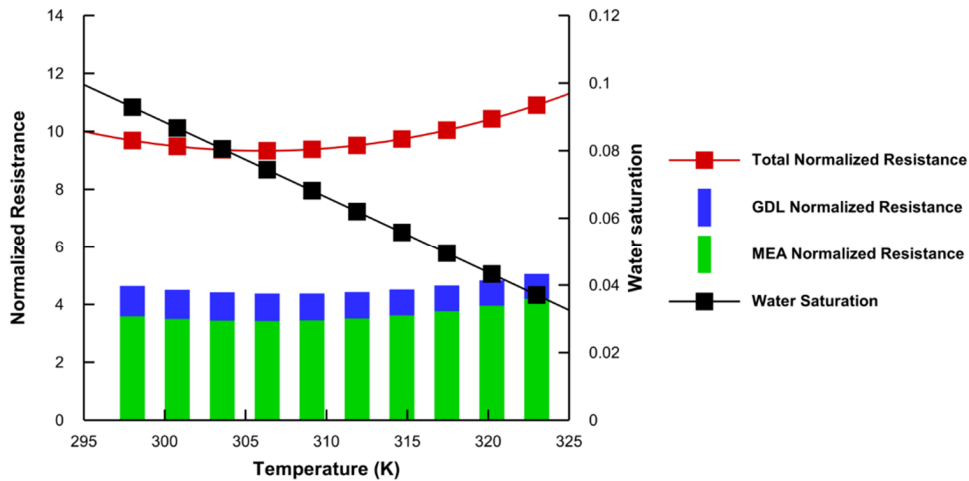

(a)

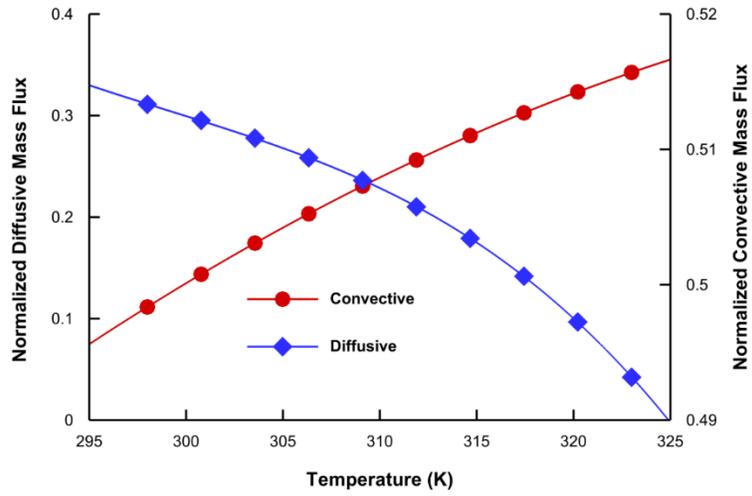

(b)

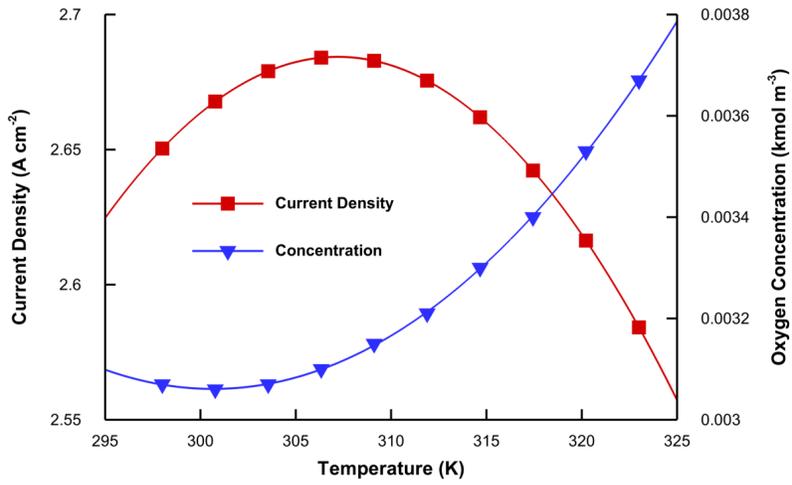

(c)

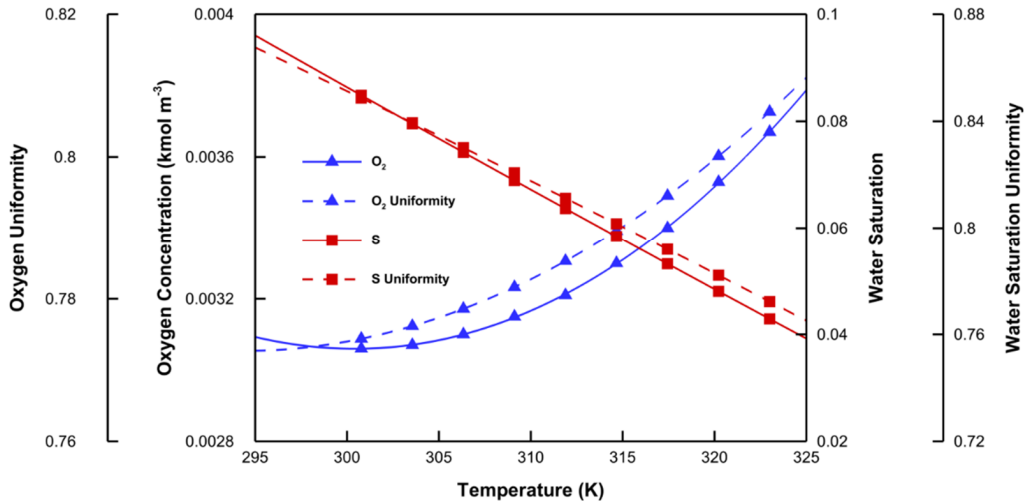

(d)

Figure 14- The effect of operating temperature on (a) normalized resistance and water saturation (b) convective and diffusive mass flux (c) current density and oxygen concentration at the reaction site (d) the distribution of reactants at the reaction site

Resistance in the flow field is essential. It contributed the most considerable portion of the total resistance in most cases. Resistance in GDL is a crucial parameter. The constant parameters were from the table of optimum features. Therefore, the resistance in GDL was low. In some cases, GDL resistance can be several times larger than MEA resistance.

7-4-Optimization

The Pareto fronts of NSGA-II and MOPSO are shown in Figure 15. The ideal point is selected from the Pareto fronts by TOPSIS, and the obtained minimum amount of objectives and optimum features are in Table 7. Results show optimum features are porosity 0.67, pressure 1 atm, temperature 323 K, anode stoichiometry 3, and cathode stoichiometry 2.62. The obtained ideal parameters are water saturation of 0.037 and the normalized resistance of 10.88.

Some references [5,35] declared the method with NSGA II has suitable runtime. NSGA-II and MOPSO reveal the same Pareto front, and NSGA runs more than 14 times slower than MOPSO. To the best of authors, the optimization time may change with the different computational systems, but the ratio does not change dramatically. MOPSO is a better MOO algorithm to predict transport phenomena in PEMFC.

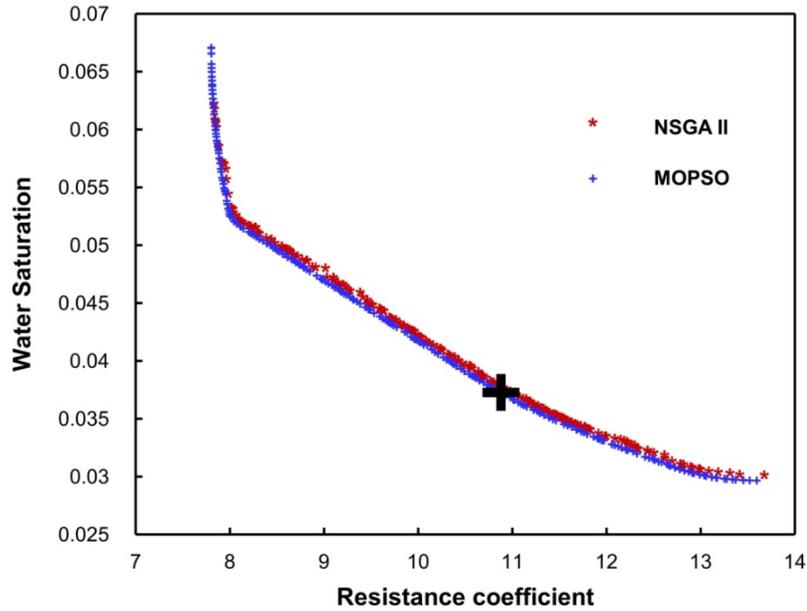

Figure 15- Pareto fronts and the ideal points

Table 7 Dimensions of components in the simulated PEMFC

| Parameter | Value |
|---|---|
| Porosity | 0.68 |
| T (K) | 323 |
| P (atm) | 1 |
| $S_a$ | 3 |
| $S_c$ | 2.62 |
| $S_{flowfield-GDL}$ | 0.037 |
| $\widehat{R}_T$ | 10.88 |

## 8- Conclusion

In this investigation, the transport phenomenon in the cathode side of an enhanced parallel flow field PEMFC was taken into account and optimized. The study procedure contains CFD and RSM for regression and MOPSO, NSGA-II, and TOPSIS for MOO. Features are the porosity of GDL, operating pressure, operating temperature, and the stoichiometric of both sides. Objectives are normalized transport resistance and water saturation at the interface of GDL and the flow field. The obtained results are:

- Two functions were developed to express oxygen transport resistance and water saturation at the highest current density of the cell, where water removal and concentration loss are significant. The models show great prediction accuracy. The $R^2$ of the resistance and water saturation models are 0.9967 and 0.974, respectively.
- Both NSGA-II and MOPSO could distinguish the Pareto front accurately, and the runtime of MOPSO was faster. Therefore, MOPSO is suggested for MOO.
- Tapering main channels can increase limiting current density by 41%.
- Pressure is the most determining feature for the resistance, and porosity is crucial for water saturation.
- By increasing pressure, the resistance in all components increases, and water saturation is lower in low pressure. The optimum operating pressure is 1 atm.

- Water saturation is higher at low cathode stoichiometry. Furthermore, the relation between cathode stoichiometry and the resistance is not linear. The optimum cathode stoichiometry is 2.62.
- A change in the stoichiometry of the anode leads to changes in water saturation and oxygen transport resistance because of the electrochemical behaviour of the cell. The best anode stoichiometry is 3.
- At high porosity, water saturation at the interface of the flow field and GDL is lower. A higher amount of resistance occurs at high porosity. The ideal porosity of GDL is 0.68.
- By increasing temperature, the water saturation decreases, and the change in the resistance is not linear. The ideal temperature is 323 K.
- The obtained optimum point is water saturation of 0.037 and normalized resistance of 10.88.

## 8- Nomenclature

| | | | |
|---|---|---|---|
| a | water activity | $\dot{N}$ | molar flux rate per unit area |
| $c_p$ | specific heat at constant pressure (J kg$^{-1}$ K$^{-1}$) | S | water saturation |
| $c_r$ | water condensation rate constant | | |
| C | molar concentration (mol m$^{-3}$) | t | time (s) |
| D | diffusion coefficient (m$^2$ s$^{-1}$) | T | Temperature (K) |
| F | Faraday constant (96487 C mol$^{-1}$) | $\vec{u}$ | velocity vector (m s$^{-1}$) |
| h | molar formation enthalpy (J mol-1) | V | voltage (V) |
| i | exchange current density (A m$^{-2}$) | $\alpha$ | transfer coefficient |
| j | volumetric exchange current density (A m$^{-3}$) | $\gamma$ | concentration dependence |
| k | thermal conductivity (W m$^{-1}$ K$^{-1}$) | $\varepsilon$ | porosity |
| K | permeability (m$^2$) | $\eta$ | overpotential (V) |
| M | molar mass (kg mol$^{-1}$) | $\theta_c$ | contact angle (°) |
| $n_d$ | electro-osmotic drag coefficient | $\lambda$ | water content in the membrane |
| P | Pressure (atm) | $\xi$ | specific active surface area (m$^{-1}$) |
| Q | mass flux (kg s$^{-1}$) | $\mu$ | dynamic viscosity (kg m$^{-1}$ s$^{-1}$) |
| $r_w$ | condensation rate (kg m$^{-2}$ s$^{-1}$) | $\rho$ | density (kg m$^{-3}$) |
| R | universal gas constant (8.314 kPa mol$^{-1}$ K$^{-1}$) | $\sigma$ | membrane conductivity (Siemens m$^{-1}$) |
| $R_T$ | transport resistance (s m$^{-1}$) | $\sigma_{surf}$ | surface tension (N m$^{-2}$) |
| RH | relative humidity (%) | $\phi$ | electric potential (V) |
| $R_{ohm}$ | ohmic resistance (ohm m) | | |

**Subscripts**

| | | | |
|---|---|---|---|
| a | anode | ocv | open-circuit voltage |
| c | cathode | r | reaction related value |
| cell | fuel cell | ref | reference value |
| g | gas phase | s | solid phase |
| $H_2$ | hydrogen | sat | Saturation |
| $H_2O$ | water | T | temperature-related value |
| i | symbol of species | u | velocity related value |
| l | liquid phase | w | water phase |
| m | membrane phase | wv | water vapor |
| $O_2$ | oxygen | | |

**Superscripts**

| | |
|---|---|
| eff | Effective |

**Acronyms**

| | | | |
|---|---|---|---|
| 3D | Three Dimensional | MOO | Multi-Objective Optimization |
| CL | Catalyst layer | MSE | Mean Square Error |
| CCD | Central Composite Design | MOPSO | Multi-Objective Particle Swarm Optimization |
| CFD | Computational Fluid Dynamic | NSGA | Non-dominated Sorting Genetic Algorithm |
| FVM | Finite Volume Method | RSM | Response Surface Method |
| GDL | Gas Diffusion Layer | PEM | Proton Exchange Membrane |
| MEA | Membrane Electrode Assembly | TOPSIS | Technique of Order Preference Similarity to the Ideal Solution |

## 9- Reference